\documentstyle[12pt,aasms4]{article}

\slugcomment{Submitted to the Astronomical Journal}
\begin{document}

\title{Ongoing Mass Transfer in the Interacting Galaxy Pair NGC 1409/10
\footnote{Based on observations with the NASA/ESA {\it Hubble Space Telescope}
obtained at the Space Telescope Science Institute, which is operated
by the Association of Universities for Research in Astronomy, Inc.,
under NASA contract No. NAS5-26555.}}

\author{William C. Keel\altaffilmark{2}}
\affil{Department of Physics and Astronomy, University of Alabama, Box 870324,
Tuscaloosa, AL 35487}

\altaffiltext{2}{Visiting Astronomer, WIYN Observatory, owned and
operated by
the WIYN Consortium, Inc., which consists of the University of Wisconsin,
Indiana University, Yale University, and the National Optical Astronomy
Observatory (NOAO). NOAO
is operated for the National Science Foundation
by the Association of Universities for Research in Astronomy (AURA), Inc.}

\begin{abstract}
I present two-band HST STIS imaging, and WIYN spectral mapping,
of ongoing mass transfer in the
interacting galaxy pair NGC 1409/10 (where NGC 1410 is the Seyfert 
galaxy also catalogued as III Zw 55). Archival snapshot WFPC2 imaging from
the survey by Malkan et al. showed a dust feature stretching between
the galaxies, apparently being captured by NGC 1409. The new
images allow estimates of the mass being transferred and rate
of transfer. An absorption lane typically 0.25" (100 pc) wide
with a representative optical depth $\tau_B = 0.2$  
cuts across the spiral structure of NGC 1410, crosses the 7-kpc projected
space between the nuclei, wraps in front of and, at the limits
of detection, behind NGC 1409, and becomes a denser ($\tau_B = 0.4$)
polar feature around the core of NGC 1409. Combination of extinction data
in two passbands allows a crude three-dimensional
recovery of the dust structure, supporting the front/back geometry
derived from colors and extinction estimates. The whole feature
contains of order $10^6$ solar masses in dust, implying about 
$2 \times 10^7$ solar masses of gas, requiring a mass transfer
rate averaging $\approx 1$ solar mass per year unless we are particularly
unlucky in viewing angle. Curiously, this demonstrable case of
mass transfer seems to be independent of the occurrence of a Seyfert 
nucleus, since the Seyfert galaxy in this pair is the donor of the material.
Likewise, the recipient shows no signs of recent star formation from
incoming gas, although NGC 1410 has numerous luminous young star clusters
and widespread H$\alpha$ emission.
\end{abstract}


\section{Introduction}

Galaxy interactions are clearly linked to bursts of star formation, and
in more restricted ways, to triggering of nuclear activity. What
physical mechanisms mediate these connections remains unclear, since
there are multiple well-motivated and plausible processes.
Among these, mass transfer between the galaxies in an interacting system has 
been long discussed, but evidence for its occurrence has remained largely
circumstantial. Such transfer could play a particular role in dumping
gas into the inner regions of a galaxy, since the angular-momentum
barrier which restricts the inward transport of disk gas may be
greatly reduced if the gas comes from another galaxy with the appropriate
encounter geometry. 
 
The conditions for mass transfer have been examined analytically as well as
numerically. Sotnikova (1990) considered the fate of dense ISM
clouds crossing between galaxies and heated by both conduction
and radiation in a hot IGM, finding that the clouds could survive
long enough to reach the second galaxy, and that a stream of clouds
from galaxies in low-eccentricity orbits could reach a quasi-steady
state configuration. Her numerical simulations (Sotnikova 1988)
indicated that this situation is reached much faster when the
spiral spin and orbit are parallel than the antiparallel case, in
roughly an orbital timescale of the binary system. 

Wallin \& Stuart (1992) presented an extensive study of mass transfer from 
restricted $n$-body simulations, including over 1000 encounter
geometries to map the systematics of mass transfer. The largest
amounts finally captured by the second galaxy are found for encounters
with the orbital plane close to the donor's disk plane, with this
favoritism enhanced by the strong dependence on orientation of
periapsis for non-planar encounters. The intuitive
increases in mass transfer for closer periapsis and larger mass
ratio also hold. These simulations do not give much detail on how the
companion captures the mass, but should be robust as to
the totals bound in its potential well. These results fit
with patterns in the 86 simulations described by Howard et al. (1993),
where the particles were tracked but there was no modelled
companion structure to interact in detail with the incoming particles.

The observational signatures of mass transfer can be hidden in the
complex kinematics and star-formation behavior of gas-rich interactions
and mergers. Thus, it has long been recognized that such effects
are easiest to see in pairs of mixed morphology, since large amounts of cool 
gas in the early-type E/S0 members of these systems very likely originated in
the gas-rich companion (as set forth by Domingue et al. 2003). This has 
been particularly discussed in the
context of emission lines, far-infrared emission, and blue (post-starburst) 
optical colors for
these galaxies, as possible evidence of past gas transfer
(de Mello et al. 1995, 1997). This would also account for the
Holmberg effect, a correlation between colors of pair
members, even for early-type members (Holmberg 1958, Demin et al. 1984,
Madore 1986, Reduzzi \& Rampazzo 1995). Marziani et al. (2003) present 
kinematic evidence of
ongoing or recent dumping of gas in the pair Arp 194, and stress
that mass transfer itself is necessary but not sufficient for cross-fuelling
of nuclear activity or starbursts.

An unusually clear example of current mass transfer appeared in the
HST WFPC2 snapshot survey of AGN and starburst
galaxies by Malkan, Gorjian, \& Tam (1998). Their
data for NGC 1409 showed a dust lane crossing between NGC 1409 and
its close companion (and Seyfert galaxy) NGC 1410, also known as
III Zw 55 (although there has been some ambuguity on the literature as
to which member of the pair is which). This feature extends even beyond
the region shown in Fig. 1 of Malkan et al. (where the galaxy
identified there as the Seyfert galaxy NGC 1410 is in fact its
earlier-type and more quiescent companion NGC 1409). Such cases
highlight the value of dust absorption as a tracer of the interstellar medium,
also exploited by several groups analyzing HST imagery of the
nuclei of early-type galaxies (Martel. et al. 2000, Tran et al. 2001,
Martini et al. 2003).

The extensive tidal envelope of stars around NGC 1409/10 provides backlighting 
to trace this feature and estimate the total mass and transfer rate involved. 
To allow such a study, two-color STIS imagery was obtained with
higher signal-to-noise in each band. The results are described here, along with
mapping in H$\alpha$ and [N II] which constrains the encounter
geometry and timing through the velocity field.

We see the NGC 1409/10 pair deep into a strong encounter, as shown by
the presence of a substantial stellar envelope extending well beyond the two 
galaxy cores, to a radius at least 15 kpc. The nuclei are projected at 
a separation of only 7 kpc, suggesting that there may have been
significant tidal loss of stars from both galaxies. The spiral NGC 1410 in 
particular
is rather small in linear size, with the nearly face-on spiral pattern 
spanning a diameter of 5.3 kpc (for $\rm H_0=72$ km s$^{-1}$ Mpc $^{-1}$
following the WMAP analysis of Spergel et al. 2003). 
This galaxy hosts a Seyfert nucleus and active disk star formation.
In contrast, NGC 1409 is a barred early-type system, best described as
type SB0. Because of the small angular extent of these structures, these 
types are not obvious from typical ground-based images (such as shown 
by Keel 1996).

\section{Observations}

\subsection{STIS imaging}

The STIS CCD imaging mode was used to observe NGC 1409/10 in
two passbands, the ``white-light" 50CCD and red F28X50LP bands. 
For study of extinction properties, the improved
surface-brightness
sensitivity compared to WFPC2, and fine angular sampling of this instrument,
outweighed the nonstandard and overlapping passbands available with its filter
complement. In this range of surface brightness, for single-orbit
exposures split in two for cosmic-ray rejection, a STIS F28X50LP
image has signal-to-noise ratio per pixel 2.5 times greater than
would be obtained with the WFPC2 PC CCD and F606W filter, a combined
result of improved sensitivity and broader passbands. Even using the 
WF chips and a pixel scale twice as coarse gives the advantage per
pixel to STIS by about 1.8 times, plus its finer sampling of the
PSF (slightly better than the Nyquist value across most of the red passband).
This improvement 
does come with accepting a smaller field of view than the entire WFPC2
outline, and of variations in the PSF across the field. From the 
cycle 13 STIS instrument handbook (Kim Quijano et al. 2003), the
fraction of encircled flux within a one-pixel radius varies by 
an extreme of 16\% around the CCD field of view, with 5\% of
this variation occurring within the region analyzed here. This
effect translates into a change in the pixel-to-pixel contrast of
the dust features, one which is small compared to the photon noise in
this application. PSF changes of similar amplitude also occur with
WFPC2; the encircled energy within a one-pixel radius varies by
5\% rms across its field. 

The very broad STIS passbands also introduce the
possibility of confusion between starlight and line emission, since
multple strong lines from photoionized material are included. This
confusion is not an important issue for these data unless there is
a close connection between the {\it ionized} gas and dust even far
from plausible ionizng sources. Beyond the Seyfert nucleus of NGC 1410,
the strongest line in equivalent width is H$\alpha$, reaching values
of 30 \AA\  in the inner disk of NGC 1410. Folding a typical set of H II 
region lines ratios
through the STIS response functions indicates that this level of line
emission would contaminate local measures of the starlight at the
3\% level in F28X50LP and less for the 50CCD mode. 
Larger contamination would require that there be substantial
emisison-line structure within the aperture size of the WIYN spectra
used for this scaling, and that such structure not be reflected in
the starlight. Since there are no potential ionizing clusters seen
near the extended dust feature, and the equivalent widths in the
extended emission occurring in this region are much smaller than
in the star-foring disk of NGC 1410, emission-line contamination is
unlikely to be an issue in these results.
 
The white-light image (in the 50CCD mode) is shown in
Fig. 1, to illustrate the geometry of the NGC 1409/10 pair, the
extensive diffuse envelope of scattered stars, and the third
companion spiral seen to the north of the strongly interacting pair.
The stellar envelope provides smooth background light to trace the
dust between and beyond the two pair members.

The observations, sequence root {\tt o5ev01}, include full-orbit
(2767 and 2815-second) sets of exposures, two in each filter
for cosmic-ray rejection.
The filter change introduces a small scale change between the
two images, evaluated from the locations of 11 unresolved or compact
objects. The angular size of the red F28X50LP pixels is larger
than the white-light 50CCD pixels by a factor $1 + 7.9 \times 10^{-4}$,
amounting to 0.8 pixel across the frame. This was corrected for
differential analysis by transforming the red image to the coordinate
system of the white-light data.

To analyze the extinction properties of the dust feature, a blue-light
image was constructed as a scaled difference between the white- and red-light
data sets. This allowed the use of the very sensitive STIS CCD for
multicolor measurements, since the exact shape of the blue passband
is less important than that this shape be known. The preflight
estimates of the relative throughputs of the two filters in their
overlap region had to be substantially revised, since using the initial
throughput values led to unphysical negative flux in the difference.
The F28X50LP filter was designed to replicate the white-light 50CCD
configuration with only a grey shift longward of
a sharp cutoff to the blue. To the extent that this was realized in
fabrication, the count rate at each pixel may be expressed
in terms of the white-light rate $C_W$ and that in the red F28X50LP
filter $C_R$ to give a differential blue rate $C_B$ according to
$$ C_W = C_B + C_R/f  \eqno{(1)}$$
where $f$ is the mean transmission ratio of red and white-light 
passbands longward of the cutoff wavelength. The in-flight calibration,
now incorporated into the STSDAS $crrefer$ files, gives
$f=0.873$ averaged across the 6000--9500 \AA\  range. This
fits with astrophysical constraints from the range of color
and direction of color gradient in the early-type galaxy, which
show that $f$ must lie in the range 0.83--0.93. Figure 2
shows the effective passbands of the two filters and their scaled difference.

An image in the blue passband was produced following equation (1).
To avoid confusion with standard photometric passbands, the original
and derived images
will be referred to simply as white-light, red, and blue.
To show the overall structure of the system, 
the central portion of the STIS field is shown in figure
3, comparing the blue and red images.

Some analysis techniques can make more appropriate use
of the observed white and red passbands, since they are
statistically independent and free of the correlated noise that
can result from measuring subtle features in a difference image.
Similarly, some points will be made more clearly using the disjoint
passbands of the red and blue images. To allow approximate scaling
of the extinction results from these broad passbands, the effective
extinction was evaluated using a model old stellar population (appropriate
for NGC 1409 and probably for most of the stellar envelope as well),
in photon units. These very broad bands have the potential for
nonlinear extinction behavior (that is, significant changes in
effective wavelength with reddening). In practice, these numerical tests
indicate that such shifts are still small for extinctions $A_V < 1$
and slopes near that of a normal Galactic extinction law ($A_V/E_{B-V}=3.1$).
The blue passband behaves very much like $V$; its extinction ranges from
$1.01-0.97 A_V$ over the range $A_V=0-4.0$, varying in a nearly linear
fashion. The red passband has extinction $0.58-0.54 A_V$ over the
same range, again changing almost linearly with $A_V$. The much
broader white-light band shows more pronounced color behavior with
increasing reddening, with extinction from $0.68-0.59 A_V$ for
$A_V=0-4.0$ in a roughly quadratic way, as the redder part of the
passband becomes more important at large extinctions. However, the
extinctions which matter in this work are small enough that these
are not major complications.

\subsection{Spectroscopic mapping}

To help understand the kinematics and interaction history of
NGC 1409/10, the velocity field in H$\alpha$ and [N II]
emission was measured using the 3.5m WIYN telescope and Dense-Pak fiber array.
As described by  Barden, Sawyer, \& Honeycutt (1998), the array
includes a $7 \times 13$ configuration of fibers in a roughly
hexagonal packing covering a $35 \times 45$" region, set for these
observations with the long axis at PA 0$^\circ$. Four outlying fibers
allow sky subtraction far from the galaxy centers.
During 6/7 December 2000, two 30-minute exposures were
obtained in each of four positions, with 60-minute exposures at
each position obtained on the following night when the transparency was 
more stable. A final 60-minute exposure was obtained with an
offset of 15" E and 16" N, to cover the fainter third galaxy just east
to NGC 1410. This galaxy, centered at (2000) $\alpha=$03h 41m 11.60s,
$\delta= -01^\circ$ 17' 40\farcs 4 from
the STIS coordinates, ended up not being well centered in any
of the fibers. The fiber location nearest its core does show
features at about the $3 \sigma$ level that match H$\alpha$ and
[N II] emission at $cz = 7808 \pm 28$ km s$^{-1}$ (heliocentric), which is 
at least
plausible given the values of 7710 and 7596 for NGC 1409 and 1410 respectively,
and the evidence that the envelope of tidally stripped stars from the
interaction encompasses this galaxy as well as the two brighter ones. 

The positions centered on NGC 1409/10 were dithered in a parallelogram pattern, 
with each leg offset by about 2", to fill the gaps between 3"
fiber apertures. The instrumental resolution with these fibers
was 1.6 \AA\ , well sampled by the 0.68-\AA\  pixel scale.
The spectral range observed was 6000-7400 \AA\ .
Velocity maps were produced from the H$\alpha$ and [N II]
$\lambda 6583$ measures; for each observation, emission was
detected in 21--28 fibers, for a total of 198 velocity measures
from the higher-quality 60-minute exposures. The
velocity maps were constructed on a 1" grid,
with overlapping aperture data averaged at each pixel and
numerical 0.1" subpixels used to track aperture outlines until the
final averaging. Registration to the direct images used reconstruction
of continuum images from the spectral measurements, giving
positions of the galaxy nuclei in the DensePak coordinate system.
The velocity field is shown in Fig. 4.

The nucleus of M32 was observed, in four exposures of 10 minutes
each in a single fiber, to provide a reference old-population spectrum.
These data are used in deriving a cross-correlation velocity
scale and in subtracting a typical bulge contribution from the
nucleus of NGC 1409.

\section{The NGC 1409/10 Encounter: Geometry, Timing and Velocity Scale}

The WIYN spectroscopic results give a velocity field (at least for ionized
gas) showing the relative velocities and senses of rotation for NGC 1409
and 1410. The northeastern side of NGC 1410 is approaching, with the
line of nodes in approximately position angle $140^\circ$. For
NGC 1409, the disk rotation is less clear, since much of the gas
may be associated with the dust lane or other tidal debris. The
velocity field in Fig. 4 shows the gradient across NGC 1409
approximately along the minor axis. The absorption-line results
from these spectra are limited, but do show a gradient across
the nucleus in the sense that the northerwestern side is receding.
For the sense of orbital motion, the H$\alpha$ results and
the central absorption-line velocity for NGC 1409 agree is
showing NGC 1409 to be receding relative to the center of mass,
with NGC 1410 approaching. The faint tails of starlight extending
beyond the STIS field (see the inset to Fig. 1) indicate the rotation
senses of the disks in the plane of the sky, since tidal
tails act as a kind of material arm and wind up with the 
sense of disk rotation. The long and prominent one-sided tail wrapping
north of NGC 1409 shows that the disk must share this sense of
rotation in projection (clockwise), fitting with a polar passage,
and putting the eastern side of the disk in front. Together, the morphology
and kinematics suggest that the orbital plane is viewed
from its eastern side, and not too nearly edge-on. The northern tidal tail also
indicates that there was a previous close passage, through the disk plane
of NGC 1409.

If the spiral pattern in NGC 1410 is trailing, the western side is
in the foreground. Only the tidal tail provides compelling evidence as to which 
orientation about the plane of the sky the disk of NGC 1409 has, since
its internal structure shows parts of a ring rather than spiral
features. NGC 1409 probably lies behind NGC 1410 from our perspective,
since the dust feature crosses in front of NGC 1409, something that
would require very special geometry if it were to be seen far from the
plane of the sky.

The observed velocity difference between nuclei is 181 km s$^{-1}$.
There are indirect arguments suggesting that we see this system
within about 30$^\circ$ of the orbital plane, so that this is
a modest underestimate of the relative velocity. NGC 1409 is
the receding member. The emission-line velocity field does
follow this trend; since dust and gas are generally associated,
the dust is likely to show this same behavior, which fits with its
apparently becoming bound to NGC 1409.

The dominant grand-design pattern in NGC 1410 is easiest to understand if
this disk is experiencing a near-planar direct (prograde) encounter,
which favors the orbital plane being viewed from its eastern side.
This would be a geometry most favorable to loss of mass from its
disk during the encounter. In contrast, NGC 1409 is undergoing a
roughly polar encounter, which would favor the acquisition of material
into a polar ring.

Fig. 5 shows a sketch of the inferred geometry of the disks and relative orbit,
incorporating the constraints from velocity field, central velocities,
and tidal structure.
To put a timescale to the encounter and mass-transfer event, the characteristic
orbital time (that is, assuming circular orbits) for the two galaxies can be 
estimated from $T = 2 \pi D / v$
for a deprojected linear separation $D$ and space-velocity difference
$v$. Including the orbital inclination to the line of sight $i$
and phase angle $\phi$ along the orbit,
measured from the plane of the sky, gives
$T= 2 D_\perp \cos i \cos \phi/ \Delta v (1-\cos^2 i \sin^2 \phi)^{1/2}$ 
in terms of
the projected linear separation $D_\perp$ and observed velocity
difference $\Delta v$ (e.g. Karachentsev 1987). The observed
$\Delta v = 181 $ km s$^{-1}$ suggests that the correction for projection 
is not large, since for even more luminous pairs, the projected
velocity differences are greater than this only 20\% of the time 
(from Karachentsev 1987). Taking the typical value of 250 km s$^{-1}$
as the maximum for galaxies at this combined luminosity (following Karachentsev,
in his table 5),
the projection effects suggest $\cos i \cos \phi > 0.72$, so that
$T = 1.7-2.7 \times 10^8$ yr. The small projected
separation of this pair, and the extensive halo of diffuse starlight,
indicate that the members are well into the spiralling stage
preceding a final merger, so that this circular-orbit timescale
is only a rough scaling value.

\section{An intergalactic dust bridge}

The NGC 1409/10 system shows a striking lane of dust, crossing between
the galaxies. The connection to the spiral pattern of NGC 1410 and
polar wrapping around NGC 1409 make it clear that the
direction of ISM transfer is from the spiral NGC 1410 to the early-type
disk of NGC 1409.

The extent and structure of the dust bridge in the NGC 1409/10 system
may be most simply shown in median-windowed images and color maps.
Fig 6 (top) shows a median-windowed image, obtained upon dividing the
original white-light image by a version median-filtered over a
2" box. From this display, the dust lane clearly crosses continuously
between the galaxies, probably connecting to a feature that crosses
the southern spiral arm of NGC 1410 and curving so as to suggest that
it passes behind most of the starlight in NGC 1409. This linkage
might then reappear as the near-polar dust structure seen closer to the
nucleus of NGC 1409. This image also highlights non-axisymmetric structure in
the smooth disk of NGC 1409, in the form of a narrow stellar bar and connected 
arcs.
The dust lane between the galaxies has only moderate observed attenuation,
with a maximum light loss in this area of 13\%. It is not completely
smooth, with multiple filaments and offsets particularly apparent as it
curves north of NGC 1409. The dust lane passes in front of the northern
part of NGC 1409, especially well  seen where the dust crosses the bar feature.
It may pass behind the southern arm of NGC 1410. The median-windowed images
in each passband were the starting point for generating maps of
residual intensity in the dust lanes, where the median-windowed
images were interpolated roughly perpendicular to the dust lanes to
make a background model, retaining the ratio of observed to model
images.

Many of the same features appear in a color map (Fig. 6), obtained as the
simple ratio of blue to red images. The signal-to-noise ratio in the dust
bridge is reduced, but structures near the center of NGC 1409 are traced
much more clearly since changes in the background illumination do not
affect the color. In front of NGC 1409 there are two distinct dust features,
one with a braided appearance wrapping across the bar, and a narrower
lane crossing almost exactly in front of the nucleus. 

To the extent that the dust properties and the distribution of starlight, 
in particular color
gradients, are known, the combination of color and intensity
information can be used to reconstruct the three-dimensional location
and intrinsic optical depth of dust features. I neglect scattering into
the observed beam at this stage, since the optical depth in the NGC 1409/10
features is modest, so that observational error dominates over this
neglect.
Let a pixel have residual intensity $R$ with respect to the unabsorbed
starlight, as determined either by simple interpolation or modelling
of the entire image structure. When a single, spatially resolved dust feature
is responsible for the flux deficit, this implies a relation between
its optical depth in the observed passband and its depth within the
starlight distribution. If $I(r)$ is the luminosity density of starlight
at location $r$ along the line of sight, the dust is located at the
position $r_0$ given implicitly by

$$ R = {{\int_{0}^{r_0} I(r) dr + e^{- \tau} {\int_{r_0}^{\infty} I(r) dr}} 
\over {\int_{0}^{\infty} I(r) dr}}   \eqno{(2)}$$

This already rules out certain locations as unphysical, formalizing the
immediate constraints that a dust feature must have optical depth at
least great enough to give the observed flux deficit, and that
it cannot lie any deeper within the starlight distribution than
an opaque absorber which gives the appropriate foreground:total
ratio. From equation (2), rearrangement shows that 
$$ \tau = \ln [{{\int_{r_0}^{\infty} I(r) dr} \over {R \int_{0}^{infty} I(r) dr
+ \int_{0}^{r_0} I(r) dr}} ] .  \eqno{(3)}$$

Data at multiple wavelengths can refine this relation further; if
the ratio of optical depths at the two wavelengths is known for
the dust, a further constraint is that the ratio of derived $\tau$
must match the reddening curve assumed for the dust, which specifies
a corresponding spatial location $r_0$ as well.

A similar approach was used by Shaya et al. (1996) in reconstructing
the locations of dust patches in NGC 1316. The most robust results,
such as front versus backside locations and derived optical depth,
do not depend on the model for the stellar luminosity distribution,
since the integral ratio gives the relative depth of the absorbing
structure into the starlight. This application may be illustrated for the
special case of no color gradient and small $\tau$, so that color terms 
due to finite passbands can be neglected. Then the
residual intensity $R_\lambda$ at each wavelength constrains the
fraction of starlight in front of the dust at its optical depth
$\tau_\lambda$ through
$$ R_\lambda = X + e^{-k \tau_\lambda} (1-X) \eqno{(4)} $$
where the extinction curve enters through $k$, the ratio of $\tau_\lambda$
to the value at a fiducial wavelength such as the $V$ band. An example
using two bands is shown in Fig. 7, where error bounds for 1\%
precision in residual intensity $R_\lambda$ are shown restricting the possible 
values of $X$ and $\tau_B$ from data in the $B$ and $R$ passbands.

Use of these expressions should be done at high enough spatial resolution to
avoid the averaging effects of dust structures within the resolution
element, which artificially flattens the extinction curve and introduces
nontrivial weighting of unresolved structures by their transmission.
WFPC2 imaging of overlapping galaxy systems shows that linear resolution
of a few tens of parsecs substantially reduces these effects
(Keel \& White 2001a,b). The STIS pixel size corresponds to
26 pc (at 105 Mpc from $\rm H_0 = 72$ km s$^{-1}$ Mpc$^{-1}$), satisfying
this condition.

For disk dust in luminous spirals, there is evidence that the
slope of the optical extinction curve is close to the typical Galactic value.
Among the four backlit spirals studied using HST imagery by
Keel \& White (2001a,b), the only one showing a significant
departure from a Galactic extinction curve is the late-type
object silhouetted in front of NGC 1275, where the unusual
environment may have altered the grain population. Dust in the
arms of NGC 2207 gives similar results from HST images, albeit with significant
errors from the less favorable geometry (Elmegreen et al. 2001).
In the nearby spiral M31, star-by-star spectroscopy indicates
that the optical behavior is much like the Milky Way (rather than,
for example, the SMC slope associated with its lower metallicity),
as found by Bianchi et al. (1996). These observations motivate the
use of
a galactic extinction law where we don't have evidence to the contrary.
However, the reddening behavior in the extended features of NGC 1409/10 
may not fit this expectation.

The multicolor comparison of reddening and extinction can, in principle, 
refine the extinction and mass
estimates of the dust bridge in NGC 1409/10. It can also suggest
whether the this bridge is spatially continuous with the other
polar features around NGC 1409, if the fading of the dust extinction
is accompanied by a greying trend which suggests that it is also
wrapping behind most of the starlight. In practice, the reddening
curve is steeper than expected for the STIS effective passbands and
a Galactic extinction law, limiting what we can learn about its spatial location
from the photometric results alone. This can be shown using the region where 
the dust passes in front of the bar of NGC 1409, and is thus plainly
in the foreground of most of the starlight (with the difference
in $X$ between here and the middle of the dust feature roughly
a factor 9 from relative starlight intensities). For a given depth of the
dust feature into the starlight and modest values of
$\tau$, changing $\tau$ for different places in the dust feature will give 
a roughly
linear locus in the color-residual intensity plane (Fig. 8).
The slope of points in this part of the dust lane is inconsistent with
``normal" reddening curves ($A_V/E_{B-V}=3.1$), implying more
reddening per unit extinction (or that the STIS passbands have
not been correctly evaluated). The slope expected for a Galactic
relation is the ratio of $k$ (as in equation 4) for the two bands,
about 0.93 in this case. A linear relation was fitted to the data from
this foreground region (region 1), incorporating the typical photometric
errors from pixel scatter in both coordinates and using the 
procedure from Press et al. (1992) as implemented in the
GSFC IDL library, giving a slope $0.800 \pm 0.019$. The
dust lane between galaxies and backlit by the diffuse envelope
light (region 2) gives a slope $0.835 \pm 0.012$. If the slope
from region 1 represents the actual dust properties for these bands,
the slope is equal to the ratio of $k$ values in the limit of
small $\tau$. For finite $X$, the locus of values of $R_1$ and
$R_2$ for changing $\tau_1$, as would be found in a single
structured dust feature, follows a family of curves whose
mean slope in a given range of $R_1$ steepens with $X$.
This happens as nonlinear structure in the $R-\tau$ relation occurs
at higher values of $R$ for dust seen deeper into the starlight.
In this case, the best fit occurs for $X=0.4 \pm 0.1$, a slight 
improvement over the intuitive conclusion that dust is most visible
for $X < 0.5$. 

This information then allows a correction to the apparent
extinction, giving typical values of $A_V$ and hence column density.  
A map of $R$ at either wavelength becomes a map of $\tau$ according to
$$e^{- \tau} = {{{R - X}}\over{{1-X}}}  \eqno{(5)} $$
(and ideally the same for each wavelength observed to within
errors of measurement and elimination of structure in the starlight).
Integration over
the resulting map of $\tau$ can give characteristic dust masses
and column densities. In this case, I follow Domingue et al. (1999)
in using a mix of graphite and silicate grain properties,
which yields a column density of $1.11 \times 10^{-10} \tau$ in
g cm$^{-2}$. The mean map for the main dust lane as converted to
$\tau_B$ was formed by averaging results from the white and red
images. Region 1 is unaffected by this scaling, being in front of
nearly all the starlight. The largest extinctions, scaled to $\tau_B$,
reach 0.55 in region 2 and 0.35 in region 1 (where the extinction is
visually more dramatic because of the brighter background intensity).
This conversion gives a dust mass of $7 \times 10^5$ solar masses in
region 2, with about $ 2 \times 10^5$ in region 1.

Additional dust features are of interest to the question of whether
the dust wraps behind NGC 1409 and reappears as the polar lanes in
one continuous structure. For region 3, where the lane appears
faintly in the median-windowed image and might be turning behind much of the
starlight, the $R_W-R_R$ relation is still steeper than for region 2.
How much steeper is poorly determined, since the fit has a slope
formally greater than unity. This would then be farther back than the
reference curve $X=0.7$ in Fig. 8. That would give this part of the lane
similar optical depth (and mass) to region 2 and support the continuity of
dust around NGC 1409. The structure in front of the central region of
NGC 1409 is in front of the bulk of the starlight ($X << 1$) by
inspection. It has $\tau_B \approx 0.4$ at its deepest points,
much like regions 1 and 2. These results are consistent with the
suspicion that there is a single continuous dust pane crossing from the
disk of NGC 1410 and wrapping in front of the northwestern edge of NGC 14109,
turning behind it, and reappearing in front on the western side near the
nucleus to form a polar structure. The total dust mass involved is about 
$2 \times 10^6$ solar masses, including the unseen portion which would
lie behind the disk of NGC 1409. For a typical gas-to-dust ratio of
160 by mass (e.g. Sodroski et al. 1994), the accompanying gas phase material 
would total $\approx 3 \times 10^8$ solar masses. The time spent crossing
along this path would be of the same order as the orbital timescale,
implying a mean mass-transfer rate of $1.1-1.4$ solar masses per year.

\section{Star formation}

The members of this pair differ as much in star formation rate as in
morphology and color. NGC 1410 shows substantial recent star
formation, including a population of luminous blue star clusters
(Figs. 1 and 6). Of these, 10 are isolated enough for simple
aperture photometry. Roughly transforming to the $V$ band gives
them observed magnitudes $22.9 - 24.4$, or absolute magnitudes
-12.2 -- -10.7. These are luminous but by no means unprecedented,
falling within the range of the brightest clusters seen in more
local interacting systems (Whitmore et al. 1999, Keel \& Borne 2003)
in which only a handful of clusters appear brighter than $M_V = -12$.
Clusters are short-lived at this luminosity (a few $10^7$ years), indicating
that their formation is ongoing at the timescale of the interaction.

Similarly, the line emission from the disk of NGC 1410 (away from the
Seyfert nucleus) indicates brisk star formation. Equivalent widths
in H$\alpha$ from 10--30 \AA\  are found throughout the region within
about 6" of the core (and excluding the active nucleus itself). This 
corresponds to a flux of $7.6 \times 10^{-14}$
erg cm$^{-2}$ s$^{-1}$, or an H$\alpha$ luminosity of $9 \times 10^{40}$
erg s$^{-1}$. Using the formulation by Kennicutt (1983) for
a Salpeter initial mass function, this corresponds to a star-formation
rate of 0.8 solar masses per year in the spiral disk of NGC 1410.

These points are in stark contrast to the lack of either associated line
emission or stellar clusters in NGC 1409 (since even the
brightest old globular clusters would be just at the threshold of these
observations). The line emission showin in Fig. 4 is kinematically
decoupled from the rotating disk of NGC 1409, and is therefore
suspect as an indicator of any UV stellar flux. The broadband color,
smooth image, and characteristic old stellar features in the spectrum
of NGC 1409 limit its overall star-formation rate to only a few per cent
of what we seen in NGC 1410. Whatever the fate of the gas reaching
NGC 1409, it is not fuelling star formation at the epoch we observe.
A star-formation rate of less than 0.05 solar mass per year would stand
out in these data.

\section{Nuclear activity}

NGC 1410 is a well-known type 2 Seyfert galaxy, and as a member of 
such a strongly-interacting pair, has been included in many samples
of AGN designed to test for links between nuclear activity and
interaction. It is therefore ironic that in this clear instance of
mass transfer, it is the Seyfert galaxy which acts as the donor, with
a much more quiescent galaxy receiving the material. Nuclear activity
in NGC 1409 is much weaker if present. The WIYN spectrum of its
nucleus is compared to NGC 1410, with and without correction for the
starlight continuum, in Fig. 9. The continuum was subtracted using
M32 as a template, rebinned to match redshifts and broadened by
a Gaussian of FWHM=11.4 \AA\  (520 km s$^{-1}$), the best fit for the 
combined velocity dispersion and central rotation gradient in the NGC 1410
spectrum. NGC 1409 shows typical LINER emission with [N II] $\lambda 6583$/H$\alpha$
near unity. Such emission, while often associated with other signs of
genuinbe nuclear activity at low luminosity, is common enough not to
be a particular indication of anything special happening in the core
of NGC1409.
 
\section{Summary}

HST STIS images in two broad passbands have been used to trace a dust
lane marking mass transfer in the interacting galaxy pair NGC 1409/10.
The combined color and extinction behavior support the impression
that there is a single feature crossing the 7-kpc gap between NGC 1410
and NGC 1409, then wrapping behind NGC 1409 and reappearing on the
opposite side to become a small lane crossing its disk over the pole.
The depth into the starlight derived from the two-color data and
the residual intensity  yield estimates of the dust mass, totalling about
$2 \times 10^6$ solar masses and likely accompanied by $3 \times 10^8$
solar masses of gas. At the characteristic orbital velocities in this
system, this suggests a mean rate of mass transfer slightly above
one solar mass per year onto NGC 1409. 

This mass-transfer calculation may be too simple,
since the feature may be more exactly described as an 
isochrone than a pipeline. It is striking that only in a small
region has material been launched away from NGC 1410, perhaps
indicating that a special location in the system was needed to
start such transfer. This fits with the fact that the dust
lane appears to cross the spiral pattern of NGC 1410.

In light of the mechanisms discussed to accont for star formation
and nuclear activity in interacting galaxies,
it is ironic that the Seyfert galaxy NGC 1410 is the donor, rather than
the recipient, of mass transfer. The recipeint NGC 1409 shows no
signs of any fate of the gas reaching its disk. Limits to its
rate of star formation stand at a few per cent of the inflow
rate, and its nuclear activity is limited to a modest LINER.
Either the flow has yet to actually reach the dense inner regions,
or some additional trigger must set in after a critical density
or mass of infalling gas is achieved. If flows such as these
drive starbursts in interacting systems, such a cycle is needed 
to drive star-formation rates an order of magnitude greater than
the estimated flow rate in NGC 1409/10.

\acknowledgments
This work was supported by NASA thought STScI grant STScI GO-8147.01-97A.
Charles Proffitt helped in resolving the STIS filter throughput issues.

\clearpage
\figcaption
{The NGC 1409/10 system, from the 50CCD white-light image. This is a 46"
section of the frame, shown with a pseudologarithmic intensity scale to
stress the diffuse and extended halo of starlight around both NGC 1409 and
1410, extending to the anonymous spiral companion to the northeast. his
halo extends well beyond the STIS frame, as shown in the inset image
taken from the Lowell 1.1m $V$ image described by Keel (1996). In particular,
there is a tail from NGC 1409 wrapping to the north of the system,
and evidence of a still fainter extension to the southwest extending
to about 125" from NGC 1410. These features indicate the net sense of
orbital motion for the pair, clockwise from our point of view and not
seen very close to the orbital plane. The white box in the inset shows the
approximate extent of the STIS image.
North is $4.6 ^\circ$ counterclockwise from the top.
\label{fig1}}

\figcaption
{The effective passbands for STIS images. This compares the throughput in
white light (S50CCD) and with the red F28X50LP filter, and the effective
blue passband from their scaled difference. To show the redward
weighting caused by counting photons from typical galaxy light,
the thin curve shows the photon spectrum of a representative
model old (13-Gyr) population
from the Charlot \& Bruzual (1991) code. The effective blue passband,
as weighted for an unreddened old population, is shown with heavier
shading.
\label{fig2}}

\figcaption
{Red and blue images of the central part of the STIS field, shown for convenience 
in the natural $x,y$ coordinates of the CCD; north is $4.6^\circ$ 
counterclockwise from left. NGC 1409 appears to the right, with the spiral
and Seyfert host NGC 1410 to the left. Each image is mapped in a 
pseudologarithmic scale starting slightly below the zero level. The connecting 
dust filament can be seen in these images, although the large dynamic
range of background light makes its entire structure 
difficult to trace in a single display. Some of the luminous star
clusters in NGC 1410 are apparent, both in the spiral arms and tracing
a circumnuclear oval.
\label{fig3}}

\figcaption
{Dense-Pak emission-line velocity field, shown as contours aligned with the
STIS white-light image. Velocities are plotted in the observed frame (before
heliocentric or galactocentric correction). The individual fibers in the
array have 3" diameter; this map results from the combination of
four dithered pointings to fill the inter-fiber gaps. The only feature
which changes when slight superresolution is used (equivalent to a
smaller pixel footprint in the drizzle algorithm) is the sharpness of the
velocity transition crossing the center of NGC 1409. The entire system
shows a single monotonic progression in radial velocity except for a small
island just above 7800 km s$^{-1}$. 
\label{fig4}}

\figcaption
{A cartoon of the orbital and rotation geometry in the NGC 1409/10 encounter,
incorporating constraints from velocity difference, the H$\alpha$+[N II]
velocity field, and morphology of tidal features. NGC 1410 is in front
and approaching, with the near side of the orbital plane to the west.
NGC 1410 is undergoing a roughly in-plane direct encounter, while the passage
experienced by NGC 1409 is nearly polar. For each disk, the thicker black edge
indicates the near side. The faint, extended tidal tails, shown
schematically by the light gray shading, both wrap in
the direction of trailing disk structure.
\label{fig5}}

\figcaption
{Dust structure as revealed in residual intensity from a median-filtered
version (top) using a 2" median window, and a color map (bottom) shown
as the ratio of white and red images. The intensity scale bars on the
right of each image run from 0.5-1.6 times the model intensity for the 
median-windowed upper
image, and from 0-2 in count-rate ratio for the bottom color image. In the lower
panel, bluer areas are white. The small red regions at each nucleus are
an artifact of the rebinning required to match these two images in scale. 
The region and orientation are the same as shown in Fig. 3. For retrieval
of the dust location and properties, the regions listed in the text are
bracketed. Region 1 is where the lane
crosses in front of the disk of NGC 1409 to the right, while region
2 is the region between the galaxies where the dust is backlit by
the diffuse stellar envelope of the binary system.
\label{fig6}}

\figcaption
{Sample retrieval of simultaneous bounds on depth $X$ of a dust region
into the starlight distribution and its optical depth $\tau_B$, as in
equation (4). For this example, the residual intensity measured
in the standard $B$ and $R$ bands is taken to have 1\% precision.
A standard Milky Way extinction law was assumed, which gives $k_B = 1$,
$k_R = 0.57$. Lines are shown for the measured $R$ and the error
bounds in each case; the allowed region from $B$ is shaded. The
error region for the combined measurements is the intersection of this
shaded region with the upper and lower error bounds from the $R$ measurement.
\label{fig7}}

\figcaption
{Measuring depth of dust features into the starlight when the pixel
signal-to-noise ratio is too low for the approach in Fig. 7 to be useful.
The points show pixels in rergion 1, silhouetted in front of the disk of 
NGC 1409, taken to define the reddening curve and thus the $X=0$
behavior. The plotted curves are for $X=0.1,0.4,0.7$ given that
reddening slope, and show the departures in the mean behavior as
fitted for region 2 between the galaxies.
\label{fig8}}

\figcaption
{Nuclear activity in NGC 1409 and 1410. The upper spectrum shows the 
observed spectrum of the nucleus of NGC1409, with the Sy 2 nucleus of
NGC 1410 scaled down by a factor 2.5 in the middle. At the bottom is
the net emission spectrum of the NGC 1409 nucleus after subtracting
a broadened M32 spectrum, to show the H$\alpha$ emission properly.
Weak LINER emission appears at $\approx$ 1-\% of the level seen in
NGC 1410.
\label{fig9}}

\end{document}